\documentclass[reprint,aps,amsmath,amssymb]{revtex4-1}
\usepackage{nccmath}
\usepackage{mathtools} 
\usepackage{graphicx}

\usepackage{dcolumn}
\usepackage{bm}
\usepackage[
margin=0.6in,
]{geometry}

\begin{document}
	
	
	\title{Instability Zones in the Dynamics of a Quantum Mechanical Quasiperiodic Parametric Oscillator}
	\author{Subhadip Biswas}
	\email{sbiswas2@sheffield.ac.uk}
	\affiliation{
		Department of Physics and Astronomy, University of Sheffield, Sheffield, S3 7RH, UK.
	}
	\author{Pratyusha Chowdhury} 
	\affiliation{
		Department of Physics, Indian Institute of Technology Guwahati, Guwahati, Assam 781039, India.
	}%
	\author{Jayanta K Bhattacharjee }
	\email{tpjkb@iacs.res.in}
	\affiliation{%
		Department of Theoretical Physics, Indian Association for the Cultivation of Science, Jadavpur, Kolkata, 700032, India.
	}%
	
	\date{\today}
	
	\begin{abstract}
		Quasi-periodically driven quantum parametric oscillators have been the subject of several recent investigations. Here we show that for such oscillators, the instability zones of the mean position and variance (alternatively the mean energy) for a time developing wave packet are identical for the strongest resonance in the three-dimensional parameter space of the quasi-periodic modulation as it is for the two-dimensional parameter space of the periodic modulations.  
		\begin{description}
			\item[Keywords]
			Quasiperiodic oscillator, Quasiperiodic \textit{Mathieu}, Quasiperiodic variance, Sub-harmonic resonance, semi classical-quantum dynamics.
			\item[pacs]
			03.65.Sq, 05.45.-a, 03.65.-w
		\end{description}
	\end{abstract}
	
	\pacs{03.65.Sq, 05.45.-a, 03.65.-w}
	\maketitle
	
	
	\renewcommand\thesection{\Roman{section}}
	\section{Introduction}
	\numberwithin{equation}{section}
	The quantum parametric oscillator has drawn a fair amount of attention because of its relevance in the physics of ion traps \cite{leibfried} as well as in the study of the longtime properties of periodically driven quantum system \cite{goldman} \cite{chu}. A particular issue about the quantum parametric oscillators is the existence of regions of instability in the plane spanned by amplitude and frequency of modulation. The mean position of an initially prepared wave packet diverges in certain regions of the plane. More importantly, the variance and hence the energy of the oscillator increases indefinitely in certain regions of the plane. The mean position diverging implies that the particle escapes from the trap. The average energy increasing indefinitely is also undesirable. It is consequently important to know the zones where these divergences occur. Recent work \cite{biswas2018}-\cite{grubelnik} has established quite clearly that the variance and the mean diverge in the same regions of the frequency-amplitude plane leading to the situation that if the particle is trapped its energy will be finite..
	
	While references \cite{biswas2018} and \cite{biswas2019} have dealt with the variance, it should be noted that reference \cite{grubelnik} has dealt with a more physical quantity which is the energy expectation value and shown that the region of divergence of mean position and the average energy are the same. We show the connection between the variance and the average energy in Sec \textbf{II}.
	
	Over the last few years, there has been a fair amount of discussion \cite{verdeny}-\cite{crowley} on quasi-periodically driven quantum systems. Some of it is related to the efforts to generalize \textit{Floquet's} theorem to quasi-periodic case  \cite{jorba}-\cite{avila}. In this work, we address the question of whether the instability regions are identical for the mean position and the variance and hence the energy for the quasiperiodically driven system as well. To answer this question, we exploit the fact that the system is a simple harmonic oscillator and hence the dynamics of the mean position and the variance (and for that matter all the higher moments) are exactly known. We look at the worst case scenario - the situation where one of the driving frequencies is in the primary parametric situation \emph{i.e.} the driving frequency is twice the natural frequency and the other frequency which makes the drive quasi-periodic is only slightly detuned leading to the most spectacular instability zones. We show numerically and through a perturbative analysis that the instability zones for the mean and variance are identical in this situation.

	The layout of the paper is as follows: In Sec \textbf{II}, we derive the dynamics of the mean position and variance of the quantum parametric oscillator for an arbitrary forcing function $f(t)$. If the natural frequency of the physical oscillator is $\omega$, then the oscillator for the mean position has the same natural frequency $\omega$ and satisfies the usual \textit{Mathieu} equation, while the oscillator describing the dynamics of the variance  will be seen to have a natural frequency of $2\omega$ (natural frequency is the frequency of the autonomous system) and is described by a third-order non-autonomous linear differential equation. For reasons which we will explain the quasiperiodic forcing function will be taken to have frequencies $\Omega$ and $\Omega (1 +\epsilon \Delta)$ where $ \epsilon << 1$. In the case the mean has its primary resonance (for $\epsilon =0$) when the response is with time period $2T = \frac{2 \pi}{\omega}$, where $T = \frac{2 \pi}{\Omega}$. The variance on the basis of \textit{Floquet's} theory would have, for $\epsilon = 0$, periodic orbits of period $3T, 2T$ and $T$ where $T = \frac{2 \pi}{2 \omega} =  \frac{ \pi}{\omega}$ is the natural period of the oscillator for the variance, which is twice the natural period of the oscillator for the mean position. It is around the periodic responses that the instability zones exist. It was shown in Ref \cite{biswas2019} from harmonic balance arguments that there is no instability zone around $T =  \frac{3 \pi}{\omega}$ and so in Sec \textbf{III} we discuss the case around $T =  \frac{2 \pi}{\omega}$, \emph{i.e.} twice the natural frequency. Here once again, a cancellation prevents the occurrence of any instability zones. We show this from a perturbative calculation which is supported by numerical calculation of the stability boundary. In Sec \textbf{IV}, we investigate the region near $T =  \frac{\pi}{\omega}$ and it is here that one finds a variety of stability zones and they agree exactly with the corresponding zones for the response near twice the natural period for the dynamics of the mean. We conclude with a brief summary in Sec \textbf{V}.

	\renewcommand\thesection{\Roman{section}}
	\section{The Equation of Motion}
	\renewcommand\thesection{\arabic{section}}
	\numberwithin{equation}{section}
	
	The one dimensional parametric oscillator is governed by the Hamiltonian  $\mathcal{H}$ given by
	\begin{eqnarray}\label{eqn:2.1}
	\mathcal{H}=\frac{p^2}{2 \mathrm{m}}+\frac{1}{2} \mathrm{m}\omega^2 x^2 (1 + \epsilon f(t)),
	\end{eqnarray}
	where $\mathrm{m}$ is the mass of the oscillator, $p$ its momentum, $x$ its coordinate and $\omega$ the frequency of the vibration. The restoring force is modified by the time dependent function $f(t)$ which has an amplitude $\epsilon$. In this work we will consider quasiperiodic $f(t)$ which can be written as,
	\begin{subequations}
	\begin{eqnarray}\label{eqn:2.2a}
	f(t) = \cos(\Omega t) + \mu \cos(1+\epsilon \Delta)\Omega t.
	\end{eqnarray}
	
	This is a special case (in many ways most interesting as we show at the end of this section) of the general form 
	\begin{eqnarray}\label{eqn:2.2b}
		f(t) = \cos(\Omega t) + \mu \cos(1+\alpha)\Omega t,
		\end{eqnarray}
	\end{subequations}
	where $\alpha$ is an irrational number.

	The \textit{Schr{\"o}dinger} equation, corresponding to the Hamiltonian of Eq.~(\ref{eqn:2.1}) is $i \hbar\dfrac{\partial \Psi}{\partial t} = \mathcal{H} \Psi$, where $\Psi(x,t)$ is the space and time dependent wave function of the system. For any operator $\mathcal{O}$ we can write down the \textit{Heisenberg} equation
	\begin{eqnarray}\label{eqn:2.3}
	i \hbar \dfrac{\hbox {d}  \mathcal{O}}{\hbox {d} t} =i \hbar\dfrac{\partial \mathcal{O}}{\partial t} + [\mathcal{O},\mathcal{H}],
	\end{eqnarray}
	where the first term on the right hand side is nonvanishing only if the operator $\mathcal{O}$ is explicitly time dependent and $[\mathcal{O},\mathcal{H}]$ is the commutator $\mathcal{O}\mathcal{H} - \mathcal{H}\mathcal{O}$. Taking the expectation value of the operators in the above equation in any state $\Psi(x,t)$, we have
	
	\begin{eqnarray}\label{eqn:2.4}
	i \hbar \dfrac{\hbox {d} }{\hbox {d} t} \langle \mathcal{O} \rangle =i \hbar \langle \dfrac{\partial \mathcal{O}}{\partial t} \rangle+ \langle [\mathcal{O},\mathcal{H}]\rangle,
	\end{eqnarray}
	where $\langle \cdots \rangle$ = $\int \hbox {d} x ~\Psi^{\ast}(x,t)(\cdots)\Psi(x,t)$. As mentioned in the introduction, our focus here will be on the expectation value of the position operator which in this case actually follows the classical equation of motion(this happens for all quadratic Hamiltonians) and the variance which is a purely quantum mechanical object and has no classical analogue except when one considers a bunch of initial conditions in classical dynamics \cite{ballentine}.
	
	The dynamics of $\langle x \rangle$, the expectation value of the position operator is found from Eq.~(\ref{eqn:2.4}) which gives
	\begin{eqnarray}\label{eqn:2.5}
	\dfrac{\hbox {d} }{\hbox {d}  t}\langle x \rangle =  \dfrac{1}{i \hbar}\langle [x,\mathcal{H}]\rangle =\frac{\langle p \rangle}{\mathrm{m}},
	\end{eqnarray}
	while,
	\begin{eqnarray}\label{eqn:2.6}
	\dfrac{\hbox {d} }{\hbox {d}  t}\langle p \rangle =  \dfrac{1}{i \hbar} \langle [p,\mathcal{H}]\rangle =- \mathrm{m}\omega^2 \langle x \rangle(1 + \epsilon f(t)).
	\end{eqnarray}
	
	The above equations lead to 
	\begin{eqnarray}\label{eqn:2.7}\nonumber
	\frac{\hbox {d}^2}{\hbox {d} t^2} \langle x \rangle + \omega^2 [1 + \epsilon \cos(\Omega t)+\epsilon \mu \cos(1+\epsilon \Delta)\Omega t ]\langle x \rangle = 0,\\
	\end{eqnarray}
	which is exactly the classical quasi-periodic \textit{Mathieu} equation (\cite{rand}; \cite{randbook}).
	
	Our primary interest here is in studying the variance and comparing its dynamics with that of the mean so far instability zones are concerned. To find the dynamics of $V$, we write
	
	\begin{subequations}
		\begin{eqnarray}\label{eqn:2.8a}
		\dfrac{\hbox {d} }{\hbox {d}  t}\langle x^2 \rangle =\dfrac{1}{2 i \hbar \mathrm{m}}\langle[x^2,p^2]\rangle=\frac{\langle xp +px \rangle}{\mathrm{m}}.
		\end{eqnarray}
		\begin{eqnarray}\label{eqn:2.8b}\nonumber
		\dfrac{\hbox {d}}{\hbox {d} t}\langle p^2 \rangle &=&\dfrac{\mathrm{m} \omega^2}{2 i \hbar }\langle[p^2,(1+\epsilon f(t))x^2]\rangle\\
		&=& -\mathrm{m} \omega^2 (1 + \epsilon f(t)) \langle xp +px \rangle .
		\end{eqnarray}
		\begin{eqnarray}\label{eqn:2.8c}\nonumber
		\dfrac{\hbox {d}}{\hbox {d} t}\langle xp +px \rangle&=&\dfrac{1}{i \hbar }\langle[xp +px, \frac{p^2}{2 \mathrm{m}}+\frac{1}{2} \mathrm{m}\omega^2 x^2 (1 + \epsilon f(t))]\rangle\\
		&=& 2\frac{\langle p^2 \rangle}{ \mathrm{m}} - 2 \mathrm{\mathrm{m}} \omega^2 \langle x^2\rangle (1 + \epsilon f(t)).
		\end{eqnarray}
	\end{subequations} 
	
	Combining the above equations, we get	
	
	\begin{eqnarray}\label{eqn:2.9}
	\begin{aligned}
	\dfrac{\hbox {d}^2}{ \hbox {d} t^2}\langle x^2 \rangle = 2\dfrac{\langle p^2 \rangle}{\mathrm{m}^2}-2 \omega^2 \langle x^2 \rangle (1 + \epsilon f(t)).
	\end{aligned}
	\end{eqnarray}
	
	Another derivative leads to 
	\begin{eqnarray}\label{eqn:2.10}
	\begin{aligned}
	\dfrac{\hbox {d}^3}{\hbox {d}t^3} \langle x^2 \rangle  =&\dfrac{2}{\mathrm{m}^2} \dfrac{\hbox {d} \langle p^2 \rangle}{\hbox {d} t}-2 \omega^2  \dfrac{\hbox {d} \langle x^2 \rangle}{\hbox {d} t} (1 + \epsilon f(t)) - 2\omega^2 \epsilon \dot{f}\langle x^2 \rangle\\
	=& - 4\omega^2(1 + \epsilon  f(t))\frac{\hbox {d}}{\hbox {d}t} \langle x^2 \rangle
	- 2\omega^2 \epsilon \dot{f}\langle x^2 \rangle.
	\end{aligned}
	\end{eqnarray} 
	
	Identical steps lead to
	\begin{eqnarray}\label{eqn:2.11}
	\dfrac{\hbox {d}^3}{\hbox {d}t^3} \langle x \rangle^2  = - 4\omega^2(1 + \epsilon  f(t))\frac{\hbox {d}}{\hbox {d}t} \langle x \rangle^2 - 2\omega^2 \epsilon \dot{f}\langle x \rangle^2,
	\end{eqnarray} 
	and hence for the variance $V = \langle x^2 \rangle -\langle x \rangle ^2$,
	
	\begin{eqnarray}\label{eqn:2.12}
	\dfrac{\hbox {d}^3 V}{\hbox {d}t^3}  + 4\omega^2(1 + \epsilon  f(t))\frac{\hbox {d} V }{\hbox {d}t} + 2\omega^2 \epsilon \dot{f} V =0.
	\end{eqnarray} 
	
	This is dynamics of the variance and with the form of $f(t)$ as given in Eq.~(\ref{eqn:2.2a}), we get
	
	\begin{eqnarray}\label{eqn:2.13}\nonumber
	\dfrac{\hbox {d}^3 V}{\hbox {d}t^3}  + 4\omega^2\frac{\hbox {d} V }{\hbox {d}t} +4 \epsilon \omega^2 [\cos(\Omega t) + \mu \cos\Omega(1+\epsilon \Delta) t]\frac{\hbox {d} V }{\hbox {d}t} \\\nonumber
	- 2\omega^2 \epsilon\Omega [\sin(\Omega t) +\mu (1+\epsilon \Delta) \sin\Omega(1+\epsilon \Delta) t] V=0.\\
	\end{eqnarray} 
	
	The fact that the behaviour of the classical system is striking was realized in the Ref~(\cite{rand}).The dynamics of the mean as shown in Eq.~(\ref{eqn:2.7}) has been extensively studied \cite{rand}-\cite{kovacic}. The dynamics of the variance as shown in Eq.~(\ref{eqn:2.13}) has been investigated in \cite{biswas2018}-\cite{biswas2019} for $\mu = 0$ (periodic modulation).  We note that from Eqs.~(\ref{eqn:2.8a})-(\ref{eqn:2.8c}), 
	
   \small{ $\frac{\hbox {d} }{\hbox {d}  t}\langle E \rangle = \frac{\hbox {d} }{\hbox {d}  t} \langle \frac{p^2}{2 \mathrm{m}}  +\frac{1}{2} \mathrm{m}\omega^2 x^2 (1 + \epsilon f(t))\rangle = -\frac{1}{2}\mathrm{m} \omega^2 \epsilon \langle x^2 \rangle\frac{\hbox {d} }{\hbox {d}  t}f(t)$}
    and hence the instability zones of $\langle E \rangle$ are linked to these of $\langle x^2 \rangle$. Our aim here will be to see how different the quasiperiodic perturbation is from the periodic one given that for the mean, the changes in the instability zone are drastic for $\mu \neq 0$ (the quasiperiodic case).
	
\begin{figure}[h!]
		\includegraphics[width= 0.48\textwidth]{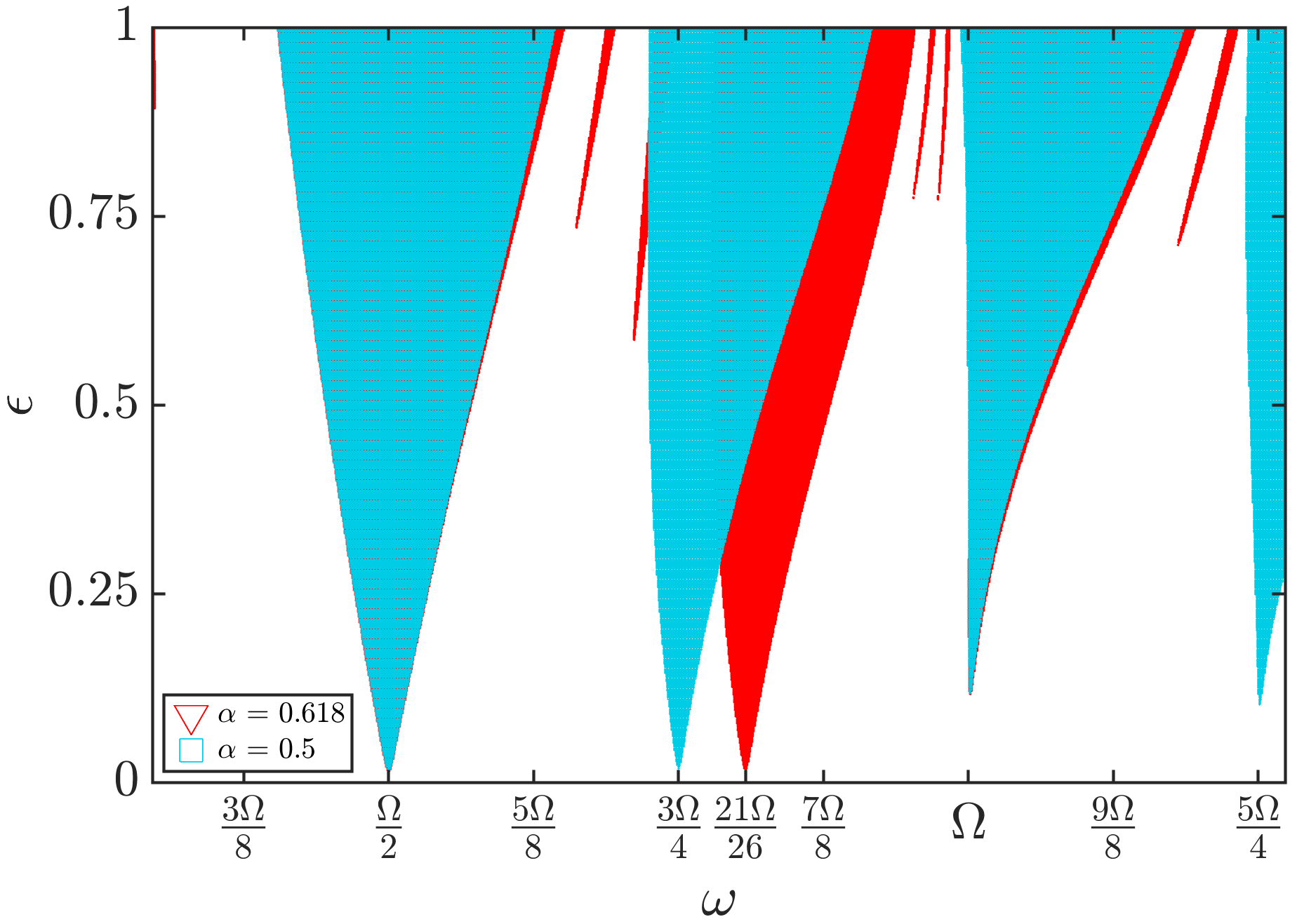}
		\caption{Numerical solution of  Eq.~(2.7) with $f(t)$ from Eq.~(\ref{eqn:2.2b}) with $\mu =0.5$, $\alpha=0.5$ and $\alpha = \frac{8}{13}$, $\Omega = 2\pi$ in the $\epsilon-\omega$ parameter space. Coloured points are the unstable solution and white region corresponds to stable solution of the \textit{Mathieu} equation. The instability zones of $V$ obtained from Eq.~(\ref{eqn:2.13}) with the same set of parameters mentioned above yield same region of the above figure.}\label{fig10}
	\end{figure}

	The technique of working with Eqs.~(\ref{eqn:2.2a}) and (\ref{eqn:2.2b}) have been carefully dealt with in Refs.~\cite{rand} - \cite{kovacic}. Here we point out the only feature that has not been explored in detail in those papers. Every irrational number has an infinite continued fraction expansion and stopping it at some point leads to a rational number approximation. What happens to the stability zones as one works with the rational approximants  and approaches the irrational number? We illustrate this limiting procedure with the golden ratio $\frac{\sqrt{5}-1}{2}$. The successive rational approximants to the irrational  $\alpha = \frac{\sqrt{5}-1}{2}$ are $\alpha = \frac{1}{2}, \frac{2}{3}, \frac{3}{5}, \frac{5}{8}, \frac{8}{13},\dots$. converging very quickly to $\alpha = 0.618 \dots$ For each of the rational approximations, $f(t)$ is a periodic function with a period $T$ that increases as the approximants converge to the irrational number. For $\alpha = \frac{1}{2}$, $f(t)$ has a period $ \frac{4 \pi}{\Omega} $, for $\alpha =\frac{2}{3}$ the period is $ \frac{6 \pi}{\Omega} $ and so on. \textit{Floquet} theory says that for $\alpha = \frac{1}{2}$, the frequencies around which one looks for instability in the response are $\frac{n \Omega}{4}$ ($n = 1, 2, 3, \dots$), while for $\alpha = \frac{2}{3}$ one looks for instability around $ \frac{n \Omega}{6} $. The basic frequency ($ \frac{\Omega}{4}, \frac{\Omega}{6}, \dots$) goes on decreasing and eventually tends to zero as $\alpha$ becomes $\frac{\sqrt{5} -1}{2}$. The fact that one is required to look for resonances near $ \frac{n \Omega}{4} $, does not mean that there will be an instability zone for all $`n'$. For $n=1$, it is straightforward to see by repeating the steps shown in Sec.~\textbf{III}, that there is no instability zone around $\frac{\Omega}{4}$. There is a periodic orbit of period $ \frac{8 \pi}{\Omega} $ (\textit{Floquet's} theorem) but it exists in the $\epsilon -\omega$ plane along the curve $\delta = -\frac{\epsilon^2}{\Omega^2}$ where $\omega^2 = \frac{\Omega^2}{16} + \delta$ for small values of $\epsilon$. The primary instability zones are around $ \frac{\Omega}{2} $ (for $n = 2$) and $ \frac{3\Omega}{4} $ (for $n = 3$), followed by $n = 4$. This picture hardly changes for the next approximant except that the significant instability zones are around $\frac{n \Omega}{6}$ with $n = 3$ and $n = 5$, followed by $n=6$. In the limiting (quasi periodic) situation the significant zones are around $\frac{n_1 \Omega}{2}$ ($ n_1  = 1, 2 $) and $n_2\frac{\sqrt{5}+1}{2}$ ($ n_2  = 1, 2 $) . The numerically obtained instability zones corroborate the above statements. The point of the present paper is that the instability zones of the variance (as obtained from the numerical or perturbative treatment of Eq.~(\ref{eqn:2.13}) ) yield the identical instability zones. We show this numerically in Fig.~(\ref{fig10}), where we have (to prevent crowding) shown the two cases of $\alpha = \frac{1}{2}$ and $\alpha=\frac{8}{13}$. The case $\alpha=\frac{8}{13}$ is virtually identical to $\alpha = 0.618$ and not very different from the first approximant $\alpha=\frac{1}{2}$. We repeat that the plots of instability zones from Eqs.~(\ref{eqn:2.7}) and (\ref{eqn:2.13}) coincide in all cases.
	
	What we see from above is that when the frequencies $\Omega$ and $\Omega(1+\epsilon \Delta)$ of Eq.~(\ref{eqn:2.2a}) are well separated, the instability zones which are most prominent when originating from $\omega = \frac{\Omega}{2}$ or $\frac{\Omega}{2}(1+\epsilon \Delta)$ are also well separated. Only if $\epsilon \Delta$ is made very small, the instability zones overlap and can produce an immense amount of fine structure as shown in Figs.~(\ref{fig1}) and (\ref{fig2}). This is what was noted by Rand \emph{et al.} \cite{rand} and consequently it is important to establish that the two immensely complicated instability zones agree exactly for the mean (Eq.~(\ref{eqn:2.7})) and variance (Eq.~(\ref{eqn:2.13})) both in a direct numerical investigation and in a perturbative
	 \textit{Krylov-Bogoliubov} analysis.  The instability zones from the \textit{Krylov-Bogoliubov} technique and direct numerical integration are shown in Sec.~\textbf{IV}.
	
	For purely periodic $f(t)$ of period $T = \frac{2 \pi}{\Omega}$ in Eq.~(\ref{eqn:2.12}), the periodic orbit can occur with periods $3T, 2T$ and $T$. This is because Eq.~(\ref{eqn:2.12}) can be written as third order traceless dynamical system having the structure $\dot{X}_{l} = \sum_{j = 1}^{3} A_{ij}(t)X_j$, where $A_{ij}(t)$ has zero trace and period $T$. We see from a harmonic balance that at lowest order there will be no instability zone around the orbit of period $3T$ and hence in the next section, we look for orbits of periodicity $2T$.

	\renewcommand\thesection{\Roman{section}}
	\section{ In the vicinity of orbits of period $2T$}
	\renewcommand\thesection{\arabic{section}}
	\numberwithin{equation}{section}

	The natural frequency of the oscillator Eq.~(\ref{eqn:2.13}) is $2\omega$ and hence if we have $2\omega$ to be in the vicinity of $\frac{\Omega}{2}$ where $\Omega = \frac{2 \pi}{T}$, we can expect a strong resonance in the system. Accordingly, we set the frequency $\omega$ near $\frac{\Omega}{4}$ and write
	
	\begin{eqnarray}\label{eqn:3.1}
	\omega = \dfrac{\Omega}{4} + \delta,
	\end{eqnarray} 
	where $\delta<<\mathcal{O}(1)$ and can be expressed in a power series in $\epsilon$ as 
	
	\begin{eqnarray}\label{eqn:3.2}
	\delta  = \delta_1 \epsilon +\delta_2 \epsilon^2 + \cdots .
	\end{eqnarray} 
	
	To $\mathcal{O}(\epsilon)$, we can now rewrite Eq.~(\ref{eqn:2.13}) as 
	
	\begin{eqnarray}\label{eqn:3.3}\nonumber
	\dfrac{\hbox {d}^3 V}{\hbox {d}t^3}  + \dfrac{\Omega^2}{4}\frac{\hbox {d} V }{\hbox {d}t}+2\Omega \epsilon \delta_1 \frac{\hbox {d} V }{\hbox {d}t}  = - \dfrac{\epsilon\Omega^2}{4} [\cos(\Omega t) +\\\nonumber
	\mu \cos\Omega(1+\epsilon \Delta) t]\frac{\hbox {d} V }{\hbox {d}t} 
	+ \dfrac{\epsilon\Omega^3}{8}[\sin(\Omega t) +\\ 
	\mu (1+\epsilon \Delta) \sin\Omega(1+\epsilon \Delta) t] V.
	\end{eqnarray} 
	
	We propose to use the \textit{Krylov-Bogoliubov} technique to explore the dynamics $V(t)$ for $\omega$ in the vicinity of $\frac{\Omega}{4}$. Accordingly, we note that for $\epsilon = 0$, the solution of Eq.~(\ref{eqn:3.3}) is
	
	\begin{eqnarray}\label{eqn:3.4}
	V_0 = A_0 + A_1 \cos\left(\dfrac{\Omega t}{2}\right) + B_1 \sin\left(\dfrac{\Omega t}{2}\right),
	\end{eqnarray} 
	where $A_0$, $A_1$ and $B_1$ are constants. To take the $\mathcal{O}(\epsilon)$ term into account, we assume that $A_0$, $A_1$ and $B_1$ will become slowly varying functions of time with the structure of the solution at the leading order unchanged. Slowly varying implies that as we take derivatives of $A_0$, $A_1$ and $B_1$, only the first order derivative will be retained, while the higher order ones will be ignored. We thus try a solution for $V(t)$ of the form (correct to $\mathcal{O}(\epsilon)$)
	
	\begin{eqnarray}\label{eqn:3.5}
	V(t) = A_0(t) + A(t) \cos\left(\dfrac{\Omega t}{2}\right) + B(t) \sin\left(\dfrac{\Omega t}{2}\right).
	\end{eqnarray} 
	
	We insert this in Eq.~(\ref{eqn:3.3}), noting that (keeping in mind that derivatives of $A_0$, $A$ and $B$ can only be of the first order)
	
	\begin{subequations}
		\begin{eqnarray}\label{eqn:3.6}\nonumber
		\dot{V}(t) = \dot{A}_0(t) + \left(\dot{A} + B \dfrac{\Omega}{2} \right) \cos\left(\dfrac{\Omega t}{2}\right) \\ 
		+  \left(\dot{B} -A\dfrac{\Omega}{2} \right) \sin\left(\dfrac{\Omega t}{2}\right).\\\nonumber
		\ddot{V}(t) =  ~~~\left(\Omega \dot{B} - A \dfrac{\Omega^2}{4} \right) \cos\left(\dfrac{\Omega t}{2}\right) ~~~~~\\
		-  \left(\Omega \dot{A}+B \dfrac{\Omega^2}{4} \right)  \sin\left(\dfrac{\Omega t}{2}\right).~~~~\\\nonumber
		\dddot{V}(t) =  -\left(\dfrac{3\Omega^2}{4} \dot{A} + B \dfrac{\Omega^3}{8} \right) \cos\left(\dfrac{\Omega t}{2}\right) \\
		-\left(\dfrac{3\Omega^2}{4} \dot{B} - A \dfrac{\Omega^3}{8} \right) \sin\left(\dfrac{\Omega t}{2}\right).
		\end{eqnarray}
	\end{subequations} 
	
	We simplify anticipating that $\dot{A}_0$, $\dot{A}$ and $\dot{B}$ will be proportional to $\epsilon$ (slowly varying implies the derivative is small) and accordingly ignore all terms which are like $\epsilon\dot{A}_0$, $\epsilon\dot{A}$ and $\epsilon\dot{B}$ \emph{etc}. Use of standerd trigonometric identities lead to Eq.~(\ref{eqn:3.3}) taking the form (correct to $ \mathcal{O}(\epsilon)$ on the right hand side) 
	
	\begin{eqnarray}\label{eqn:3.7}\nonumber
	\dfrac{\Omega^2}{4} \dot{A}_0 - \dfrac{\Omega^2}{2} \dot{A}  \cos\left(\dfrac{\Omega t}{2}\right)- \dfrac{\Omega^2}{2} \dot{B}  \sin\left(\dfrac{\Omega t}{2}\right)~~~~~~~~~~~~~~~~~~~~\\\nonumber
	=\epsilon \Omega^2 \delta_1 \left[ A\sin\left(\dfrac{\Omega t}{2}\right) - B\cos\left(\dfrac{\Omega t}{2}\right)\right]\\\nonumber
	+\frac{\epsilon \Omega^3}{8} \left[A_0 \sin(\Omega t) + A\sin\left(\dfrac{3\Omega t}{2}\right) -B \cos\left(\dfrac{3\Omega t}{2}\right)\right.\\\nonumber\left.+ \mu A\sin\left(\dfrac{3\Omega }{2} + \epsilon \Omega \Delta\right)t -\mu B \cos\left(\dfrac{3\Omega }{2}+ \epsilon \Omega \Delta\right)t\right].\\
	\end{eqnarray} 
	
	Matching the coefficients of similar trigonometric terms on the left and right sides of Eq.~(\ref{eqn:3.5}), we have
	
	\begin{subequations}
		\begin{eqnarray}\label{eqn:3.8}
		\dot{A}_0 &=& ~~~~0.\\                                  
		\dot{A} ~~&=& ~~~2\delta_1 B \epsilon.\\
		\dot{B}~~ &=& -2\delta_1 A \epsilon.
		\end{eqnarray}
	\end{subequations}

	If we look at the $\omega-\epsilon$ plane, then starting at the point at $\omega = \frac{\Omega}{4}$, the trajectory is  periodic along the vertical line. Starting form the $\omega = \frac{\Omega}{4}$ and if we move away a small distance $\delta_1$ from $\omega = \frac{\Omega}{4}$, the trajectory is quasi-periodic with the frequencies $\frac{\Omega}{2} \pm 2 \delta_1 \epsilon$. Thus there is no instability zone of Eq.~(\ref{eqn:2.13}) around $\omega = \frac{\Omega}{4}$ to $\mathcal{O}(\epsilon)$. This is exactly what had happened for the purely periodic case of $\mu = 0 $ and no qualitative change occurs for $\mu \neq 0$.

	\renewcommand\thesection{\Roman{section}}
	\section{ In the vicinity of orbits of period $T$}
	\renewcommand\thesection{\arabic{section}}
	\numberwithin{equation}{section}
	
	In this case, our natural frequency $2\omega$ in Eq.~(\ref{eqn:2.13}) needs to be close to the forcing frequency $\Omega$ and hence
	
	\begin{eqnarray}\label{eqn:4.1}
	\omega = \dfrac{\Omega}{2} +\delta =  \dfrac{\Omega}{2} +\epsilon \delta_1 + \mathcal{O}(\epsilon^2).
	\end{eqnarray} 
	
	We first recall the results for the mean position (Eq.~(\ref{eqn:2.7})), where we try the solution (\textit{Krylov-Bogoliubov}) 
	
	\begin{eqnarray}\label{eqn:4.2}
	\langle x \rangle = A(t) \cos \left( \dfrac{\Omega t}{2} \right) + B(t) \sin \left( \dfrac{\Omega t}{2} \right),
	\end{eqnarray} 
	where $A(t)$ and $B(t)$ are slowly varying amplitudes. They are easily seen to have the dynamics (with $\tau = \Delta \epsilon t \Omega$)
	\begin{subequations}
		\begin{eqnarray}\label{eqn:4.3a}
		\Delta \dfrac{\hbox {d} A}{\hbox {d} \tau} &=&~~~ \left(\dfrac{\delta_1}{\Omega} - \dfrac{1}{8} \right)B - \dfrac{\mu A}{8} \sin \tau  - \dfrac{\mu B}{8} \cos \tau .\\\label{eqn:4.3b}
		\Delta \dfrac{\hbox {d} B}{\hbox {d} \tau} &=& -\left(\dfrac{\delta_1}{\Omega} +\dfrac{1}{8} \right)A - \dfrac{\mu A}{8} \cos \tau  + \dfrac{\mu B}{8} \sin \tau .
		\end{eqnarray}
	\end{subequations} 
	\begin{figure}[ht]
		\includegraphics[width= 0.49\textwidth]{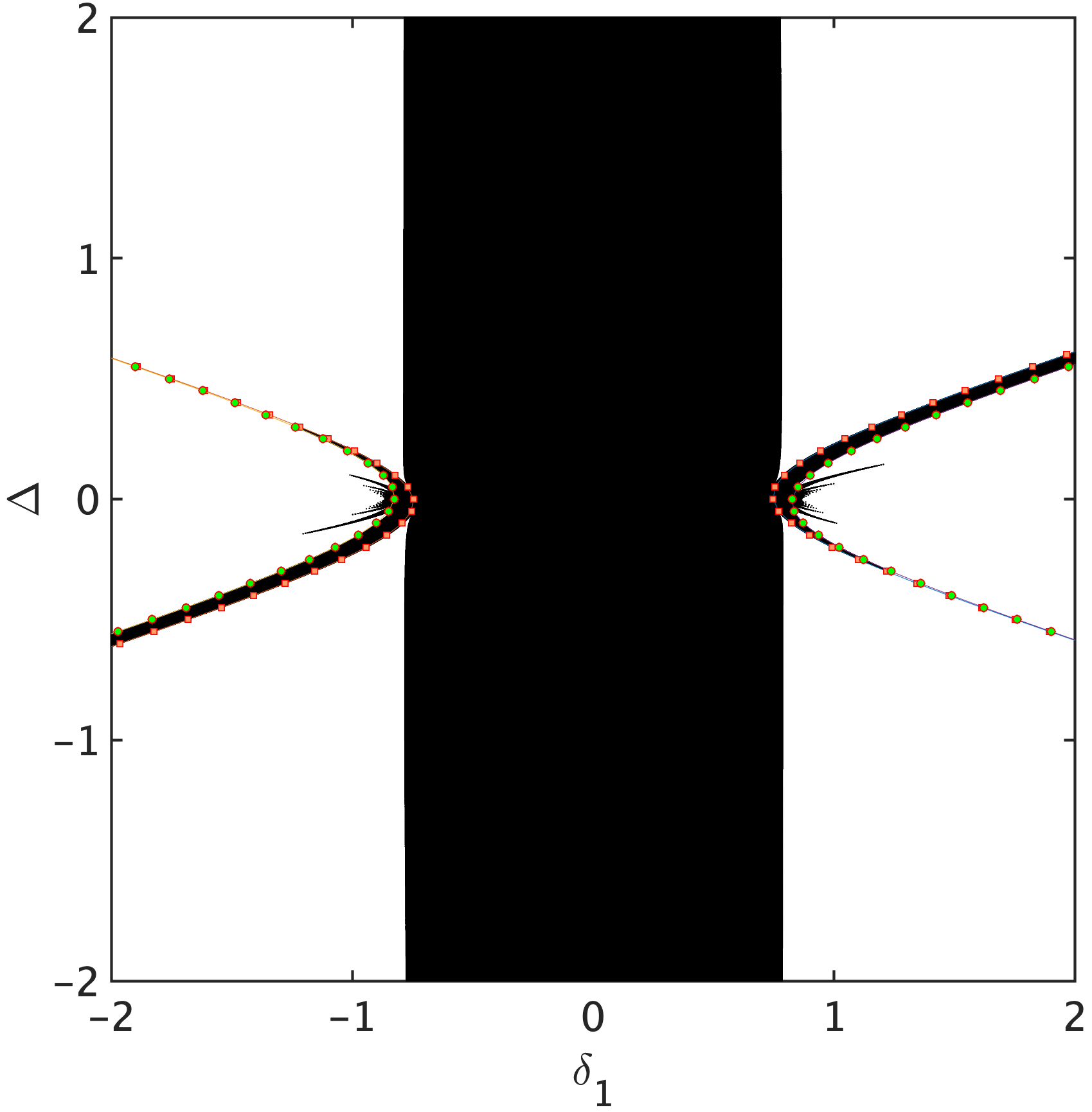}
		\caption{Zones of stable and unstable region obtained by numerically integrating the slow flow Eqs.~(\ref{eqn:4.3a}) and (\ref{eqn:4.3b}) for $\mu = 0.1$. Black points correspond to unstable regions, whereas white regions are stable. The instability zones of $V$ obtained from Eqs.~(\ref{eqn:4.12a}) - (\ref{eqn:4.12c}) for $\mu =0.1$ agree with black region of the above figure. Dotted lines are plotted from the analytical solution of Eq.~(\ref{eqn:4.6}) which has been discussed later.}
		\label{fig1}
	\end{figure}

\begin{figure}[h!]
		\includegraphics[width= \textwidth/2]{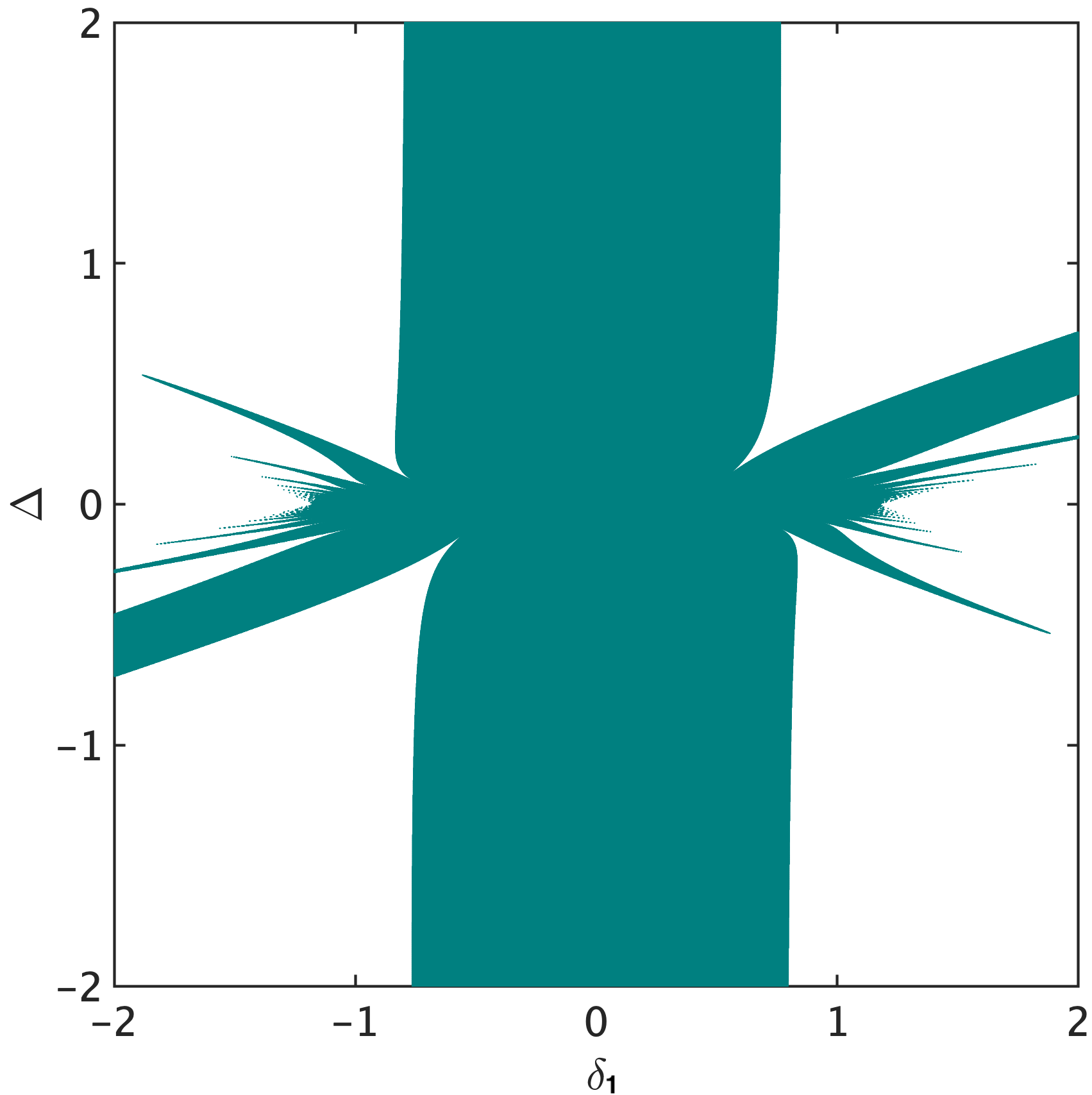}
		\caption{Zones of stable and unstable region obtained by numerically integrating the slow flow Eqs.~(\ref{eqn:4.3a}) and (\textit{4.3b}) for $\mu = 0.5$. \emph{Teal} coloured points correspond to unstable regions, whereas white regions are stable. The instability zones of $V$ obtained from Eqs.~(\ref{eqn:4.12a}) - (\ref{eqn:4.12c}) for $\mu =0.5$ agree with coloured region of the above figure.}
		\label{fig2}
	\end{figure}

\begin{figure}[h!]
		\includegraphics[width= \textwidth/2]{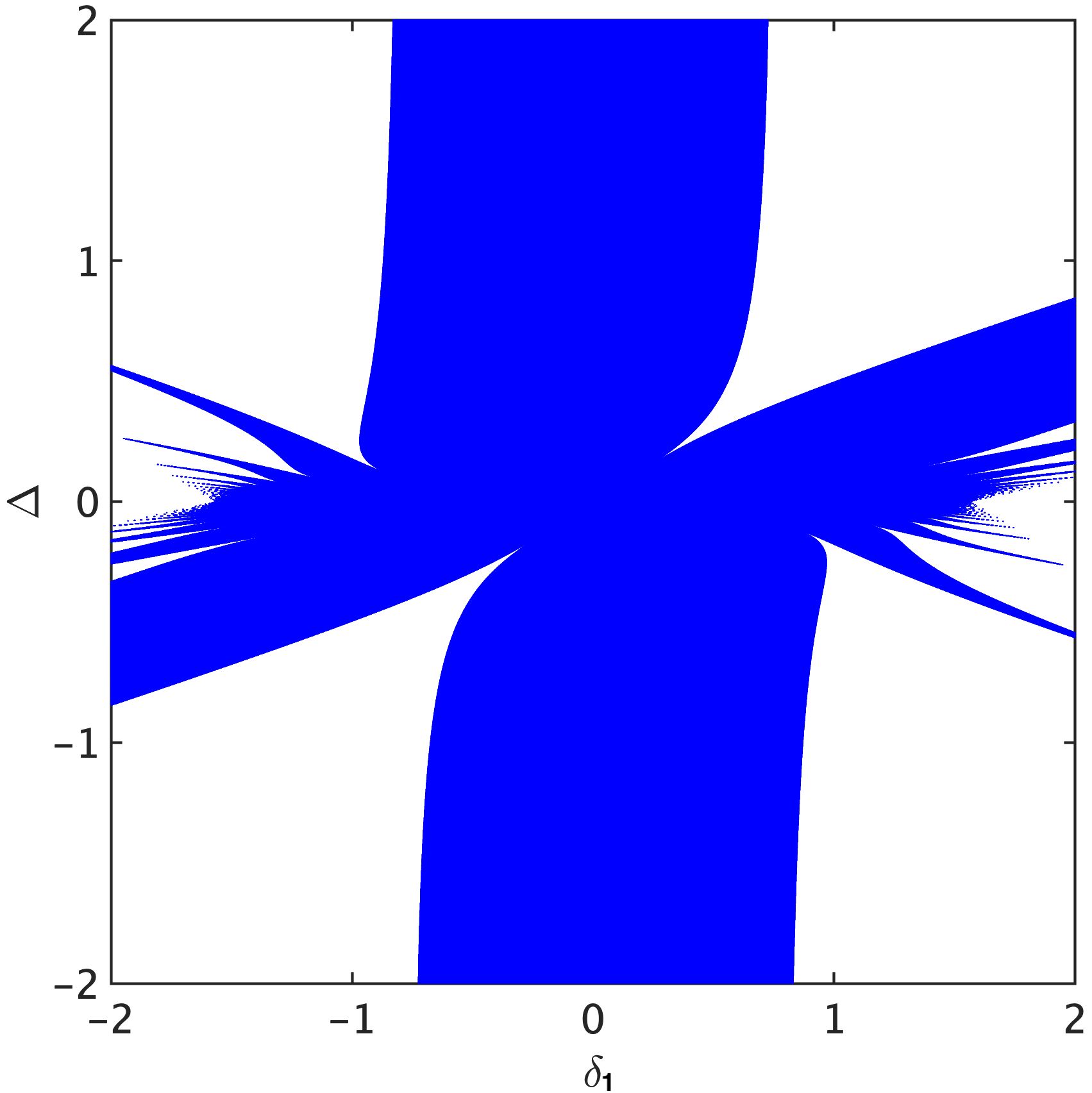}
		\caption{Zones of stable and unstable region obtained by numerically integrating the slow flow Eqs.~(\ref{eqn:4.3a}) and (\textit{4.3b}) for $\mu = 1$. Blue points correspond to unstable regions, whereas white regions are stable. The instability zones of $V$ obtained from Eqs.~(\ref{eqn:4.12a}) - (\ref{eqn:4.12c}) for $\mu =1$ agree with blue region of the above figure.}
		\label{fig3}
	\end{figure}

\begin{figure}[h!]
	\includegraphics[width= \textwidth/2]{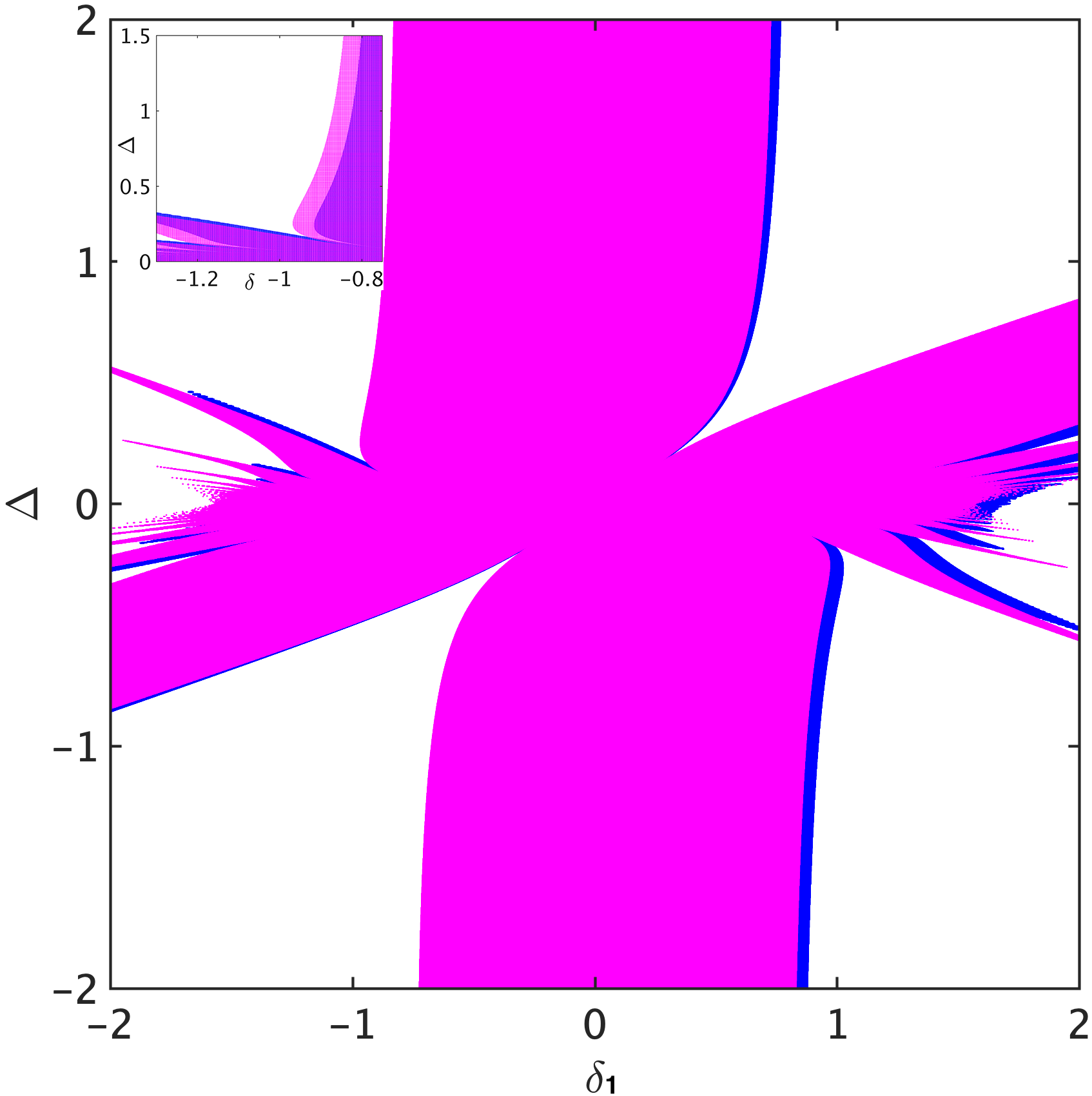}
	\caption{Numerical solution of Eqs.~(\ref{eqn:4.3a})-(\ref{eqn:4.3b}) and Eq.~(\ref{eqn:2.7}) with $\mu =1$, $\epsilon =0.1$, $\Omega = 2\pi$ in the $\Delta-\delta$ parametric space. Magenta points are the unstable solution of slow-flow of \textit{Mathieu} equation \emph{i.e.} Eqs.~(\ref{eqn:4.3a})-(\ref{eqn:4.3b}) and blue points are the unstable solution of \textit{Mathieu} equation \emph{i.e.} Eq.~(\ref{eqn:2.7}). Inset is a blowup of $\delta$ from -1.2 to -0.8 and $\Delta$ from 0 to 1.5.}\label{fig4}
\end{figure}

\begin{figure}[h!]
		\includegraphics[width= \textwidth/2]{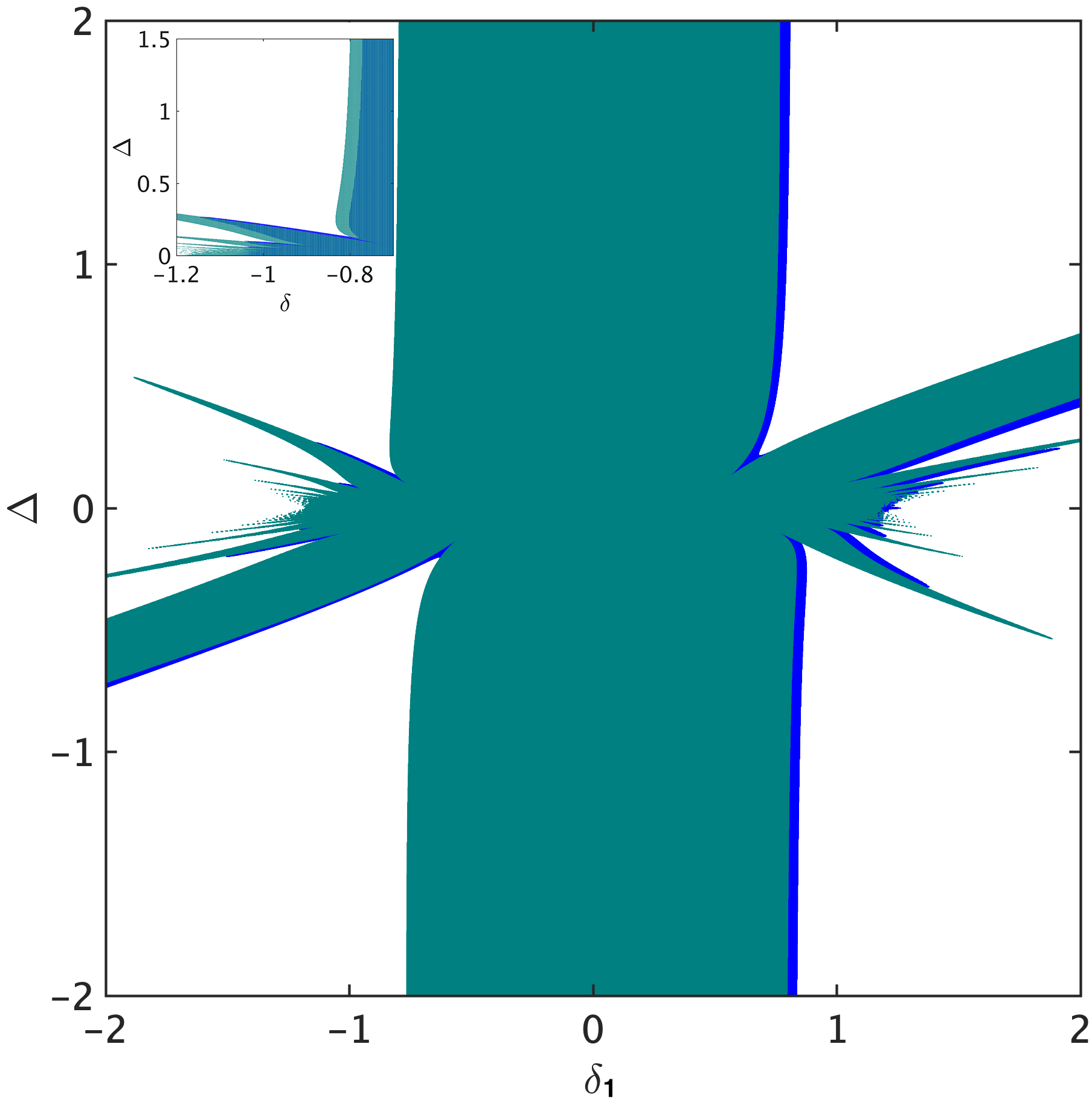}
		\caption{Numerical solution of Eqs.~(\ref{eqn:4.12a})-(\ref{eqn:4.12c}) and Eq.~(\ref{eqn:2.13}) with $\mu =0.5$, $\epsilon =0.1$, $\Omega = 2\pi$ in the $\Delta-\delta$ parametric space. \emph{Teal} coloured points correspond to the unstable solution of slow-flow of \textit{variance} equation \emph{i.e.} Eqs.~(\ref{eqn:4.12a})-(\ref{eqn:4.12c}) and blue points are the unstable solution of \textit{variance} equation \emph{i.e.} Eq.~(\ref{eqn:2.13}). Inset is a blowup of $\delta$ from -1.2 to -0.8 and $\Delta$ from 0 to 1.5.}\label{fig5}
	\end{figure}

	These flows agree exactly with the results following from the two-time scale technique of Rand \emph{et al.} \cite{rand}. For $\mu = 0$, the results are in exact agreement with those known for the \textit{Mathieu} equation. For $\mu \neq 0 $, the divergence zone changes. The results in the case are first presented numerically. We show the result with $\delta_1$ along the $x-$axis and $\Delta$ (the quasi-periodicity causing detuning parameter) along the $y-$axis. Each plot corresponds to different value of $\mu$. We have considered three values of $\mu$, namely $\mu =0.1,~0.5$ and $1$. The instability zones are shown in Figs.~(\ref{fig1}),(\ref{fig2}) and (\ref{fig3}). A comparison with the instability zones coming from the exact Eq.~(\ref{eqn:2.7}) and Eq.~(\ref{eqn:2.13}) is shown in Fig.~(\ref{fig4}) and Fig.~(\ref{fig5}) respectively. It should be noted that Figs.~(\ref{fig1})-(\ref{fig3}) correspond to the perturbation theory results and are independent of $ \epsilon $ and for different values of $ \mu $. What these figures establish is that the instability zones of the mean and variance coincide in perturbation theory. In Figs.~(\ref{fig4}) - (\ref{fig5}) we show how the full equation when plotted for definite but small $ \epsilon $ show small deviations from the approximation but the result that the instability zones coincide hold.


	We have used \textit{Runge-Kutta} method to solve two first order coupled non-autonomous differential equations. Discretization of first order ODE has the form $y_{i+1}-y_{i} = h \phi (y_i,x_i,h)$, where $h$ is the step size and $\phi$ is the increment function of $f(x,y)$ in the interval $x_{i+1} \geq x \geq x_i$. Initially, at $\tau = 0$ we choose the initial value of $\large{y}_{\tau = 0}$ as $ 0.001 $. We are looking at the values of these functions at $\tau = 7000$ with step size $h=0.001$. Stable oscillatory solution gives with amplitudes with $\sim 10^{-2}$ in the white region, whereas, functional values $\large{y}_{\tau} \geq 10^{1}$ corresponds to divergent solution. Numerical solutions of $\large{y}_{\tau}$ is shown in the Figs.~(\ref{fig8}) - (\ref{fig9}). All the values of the initial conditions that have used to solve the equations are $0.001$.  Other initial values and $h$ do not affect the stability chart diagram.  

	We would like to understand the existence of instability zones from a perturbation theory approach for $\mu < < 1$ and then compare with an exact numerical integration of Eq.~(\ref{eqn:2.7}). By an inspection of Eqs.~(\ref{eqn:4.3a})-(\ref{eqn:4.3b}) from a self consistency perspective, we try out a solution of the form 
	
	\begin{subequations}
		\begin{eqnarray}\label{eqn:4.4}
		A(\tau) &=& \alpha_1 \cos \left(\dfrac{\tau}{2}\right) + \beta_1 \sin\left(\dfrac{\tau}{2}\right) .\\
		B(\tau) &=& \alpha_2 \cos \left(\dfrac{\tau}{2}\right) + \beta_2 \sin\left(\dfrac{\tau}{2}\right) .
		\end{eqnarray}
	\end{subequations}

	where $\alpha_1$ , $\beta_1$,$\alpha_2$ and $\beta_2$ are functions of the time variable $\tau$. Inserting the above in Eqs.~(\ref{eqn:4.3a})-(\ref{eqn:4.3b}) and equating coefficients of $\cos\left(\frac{\tau}{2}\right) $ and $\sin\left(\frac{\tau}{2}\right) $, we get 
	\begin{fleqn}[\parindent]
		\begin{subequations}
			\begin{eqnarray}\label{eqn:4.5}
			\Delta \dot{\alpha}_1 + \left(\dfrac{\Delta}{2} + \dfrac{\mu}{16} \right) \beta_1 - \left[ \dfrac{\delta_1}{\Omega} - \dfrac{1}{8} \left(1+\dfrac{\mu}{2} \right)\right] \alpha_2 = 0.\\
			\Delta \dot{\beta}_1 - \left(\dfrac{\Delta}{2} - \dfrac{\mu}{16} \right) \alpha_1 - \left[ \dfrac{\delta_1}{\Omega} - \dfrac{1}{8} \left(1-\dfrac{\mu}{2} \right)\right] \beta_2 = 0.\\
			\Delta \dot{\alpha}_2 + \left(\dfrac{\Delta}{2} - \dfrac{\mu}{16} \right) \beta_2 + \left[ \dfrac{\delta_1}{\Omega} + \dfrac{1}{8} \left(1+\dfrac{\mu}{2} \right)\right] \alpha_1 = 0.\\
			\Delta \dot{\beta}_2 - \left(\dfrac{\Delta}{2} + \dfrac{\mu}{16} \right) \alpha_2 + \left[ \dfrac{\delta_1}{\Omega} - \dfrac{1}{8} \left(1-\dfrac{\mu}{2} \right)\right] \beta_1 = 0.
			\end{eqnarray}
		\end{subequations} 
	\end{fleqn}
	The above set of equations have the form $\dot{X}_i = A_{ij} X_j$.
	The border between stability and instability is obtained from the condition that matrix A has a zero eigenvalue. This gives the conditions 
	
	\begin{equation}\label{eqn:4.6}
	\dfrac{\delta_1}{\Omega} = 
	\begin{dcases} 
	-\dfrac{1}{16} \left[\sqrt{(\mu - 8\Delta)^2 +4} +\mu \right], \\
	~~\dfrac{1}{16} \left[\sqrt{(\mu - 8\Delta)^2 +4} -\mu \right], \\
	-\dfrac{1}{16} \left[\sqrt{(\mu + 8\Delta)^2 +4} -\mu \right], \\
	~~\dfrac{1}{16} \left[\sqrt{(\mu + 8\Delta)^2 +4} +\mu \right].
	\end{dcases}
	\end{equation}

	The above boundaries for $\mu=0.1$ are the same as the exact numerical results shown in Fig.~(\ref{fig1}), whereas for larger values of $\mu$, boundaries are not exactly the same as described in Eq.~(\ref{eqn:4.6}).
	
	We now turn to the dynamics of $V$ and inserting Eq.~(\ref{eqn:4.1}) for the frequency $\omega$ in Eq.~(\ref{eqn:2.13}) , obtain
	
	\begin{eqnarray}\label{eqn:4.7}\nonumber
	\dfrac{\hbox {d}^3 V}{\hbox {d}t^3}  + \Omega^2\frac{\hbox {d} V }{\hbox {d}t}+4\Omega \epsilon \delta_1 \frac{\hbox {d} V }{\hbox {d}t}  = - \epsilon\Omega^2 [\cos(\Omega t) +\\\nonumber
	\mu \cos\Omega(1+\epsilon \Delta) t]\frac{\hbox {d} V }{\hbox {d}t} 
	+ \dfrac{\epsilon\Omega^3}{2}[\sin(\Omega t) +\\ 
	\mu (1+\epsilon \Delta) \sin\Omega(1+\epsilon \Delta) t] V.
	\end{eqnarray} 

	We will approach these again in the manner of sec.~(\textbf{III}) and for $\epsilon = 0$, we write
	
	\begin{eqnarray}\label{eqn:4.8}
	V_0 = A +  B \cos(\Omega t)+ C\sin(\Omega t),
	\end{eqnarray} 
	
	where $A,B$ and $C$ are constants. In the \textit{Krylov-Bogoliubov} spirit, we now make $A,B$ and $C$ slowly varying in time (\emph{i.e.} $\dot{A},\dot{B}$ and $\dot{C}$ are of $\mathcal{O}(\epsilon)$) and proceeding exactly as in the previous section, we have correct to lowest order in $\epsilon$, for terms on the left hand side of Eq.~(\ref{eqn:4.7}) 
	
	\begin{subequations}
		\begin{eqnarray}\label{eqn:4.9}\nonumber
		\dddot{V}&=&-3\Omega^2 \dot{A} \cos(\Omega t) - 3\Omega^2 \dot{B} \sin(\Omega t) \\
		&~&+ \Omega^3 A \sin(\Omega t)- \Omega^3 B \cos(\Omega t).\\\nonumber
		\Omega^2 \dot{V} &=& \Omega^2 \dot{C} + \Omega^2 \dot{A} \cos(\Omega t)+ \Omega^2 \dot{B} \sin(\Omega t)\\&~&- \Omega^3 A \sin(\Omega t)+ \Omega^3 B\cos(\Omega t).\\
		4\Omega \delta_1 \epsilon \dot{V} &=& 4\Omega \delta_1 \epsilon (- \Omega A \sin(\Omega t)+ \Omega B\cos(\Omega t)).
		\end{eqnarray}
	\end{subequations}

	
	This yields
	\begin{eqnarray}\label{eqn:4.10}\nonumber
	\dddot{V}  + \Omega^2 \dot{V}+4\Omega \epsilon \delta_1 \dot{V}  = \Omega^2 \dot{C} -2 \Omega^2 \dot{A} \cos(\Omega t)\\\nonumber
	-2 \Omega^2 \dot{B} \sin(\Omega t)-4 \Omega \delta_1 A \sin(\Omega t) +4 \Omega \delta_1 B \cos(\Omega t).\\
	\end{eqnarray} 
	
	For the \textit{r.h.s.} of Eq.~(\ref{eqn:4.2}) evaluation to $\mathcal{O}(\epsilon)$ simply requires using $V_0$ and $\dot{V}_0$ in the \textit{r.h.s.}  of Eq.~(\ref{eqn:4.2}) and keeping the terms which have the same structure as on the \textit{r.h.s.} of Eq.~(\ref{eqn:4.3a}). We finally arrive at
	
	\begin{eqnarray}\label{eqn:4.11}\nonumber
	\dot{C}&-&2 \dot{A} \cos(\Omega t) -2 \dot{B} \sin(\Omega t) -  \dfrac{4 \delta_1}{\Omega }A \sin(\Omega t)\\\nonumber
	&+&  \dfrac{4 \delta_1}{\Omega } B\cos(\Omega t) + \dfrac{\epsilon\Omega }{4} \left[ B+ \mu A \sin(\epsilon \Delta \Omega t) \right.\\\nonumber 
	&+&\mu B \cos(\epsilon \Delta \Omega t)-2\mu C \sin( \Omega t) \left(1 + \mu \cos(\epsilon \Delta \Omega t) \right)\\
	&-&2\mu C \cos(\Omega t)\sin(\epsilon \Delta \Omega t)\left.\right].
	\end{eqnarray} 
	
	Defining $\tau = \epsilon \Delta \Omega t$, we have correct to $\mathcal{O}(\epsilon)$
	
	\begin{subequations}
		\begin{eqnarray}\label{eqn:4.12a}
		\Delta \dfrac{\hbox {d} C}{\hbox {d} \tau} &=& - \dfrac{B}{4} - \dfrac{\mu A}{4} \sin(\tau)- \dfrac{\mu A}{4} \cos(\tau).\\\label{eqn:4.12b}
		\Delta \dfrac{\hbox {d} A}{\hbox {d} \tau} &=& ~~~\dfrac{2 \delta_1 B}{\Omega} - \dfrac{\mu C}{4} \sin(\tau).\\\label{eqn:4.12c}
		\Delta \dfrac{\hbox {d} B}{\hbox {d} \tau} &=&  -\dfrac{2 \delta_1 A}{\Omega}- \dfrac{\mu C}{4} \cos(\tau)- \dfrac{C}{4} . 
		\end{eqnarray}
	\end{subequations}  
	
	\begin{figure*}
		\centering
		\begin{minipage}[t]{0.48\textwidth}
			\centering
			\includegraphics[width= \textwidth]{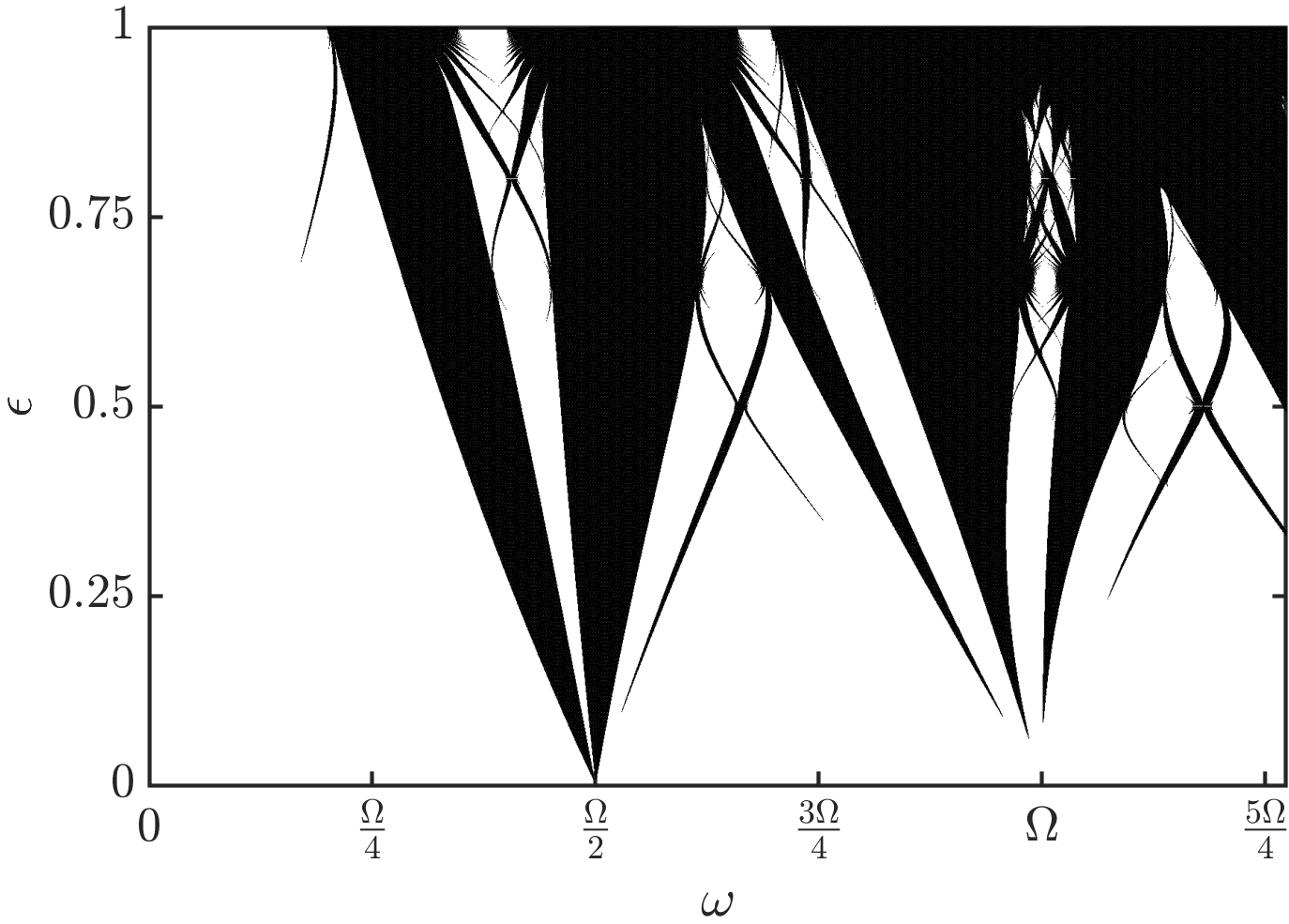}
			\caption{Numerical solution of  Eq.~(2.7) with $\mu =1$, $\Delta=-0.5$, $\Omega = 2\pi$ in the $\epsilon-\omega$ parametric space. Black points are the unstable solution and white region corresponds to stable solution of the \textit{Mathieu} equation. The instability zones of $V$ obtained from Eq.~(\ref{eqn:2.13}) with the same set of parameters mentioned above yield same region of the above figure.}\label{fig6}
		\end{minipage}\quad
		\begin{minipage}[t]{0.48\textwidth}
			\includegraphics[width= \textwidth]{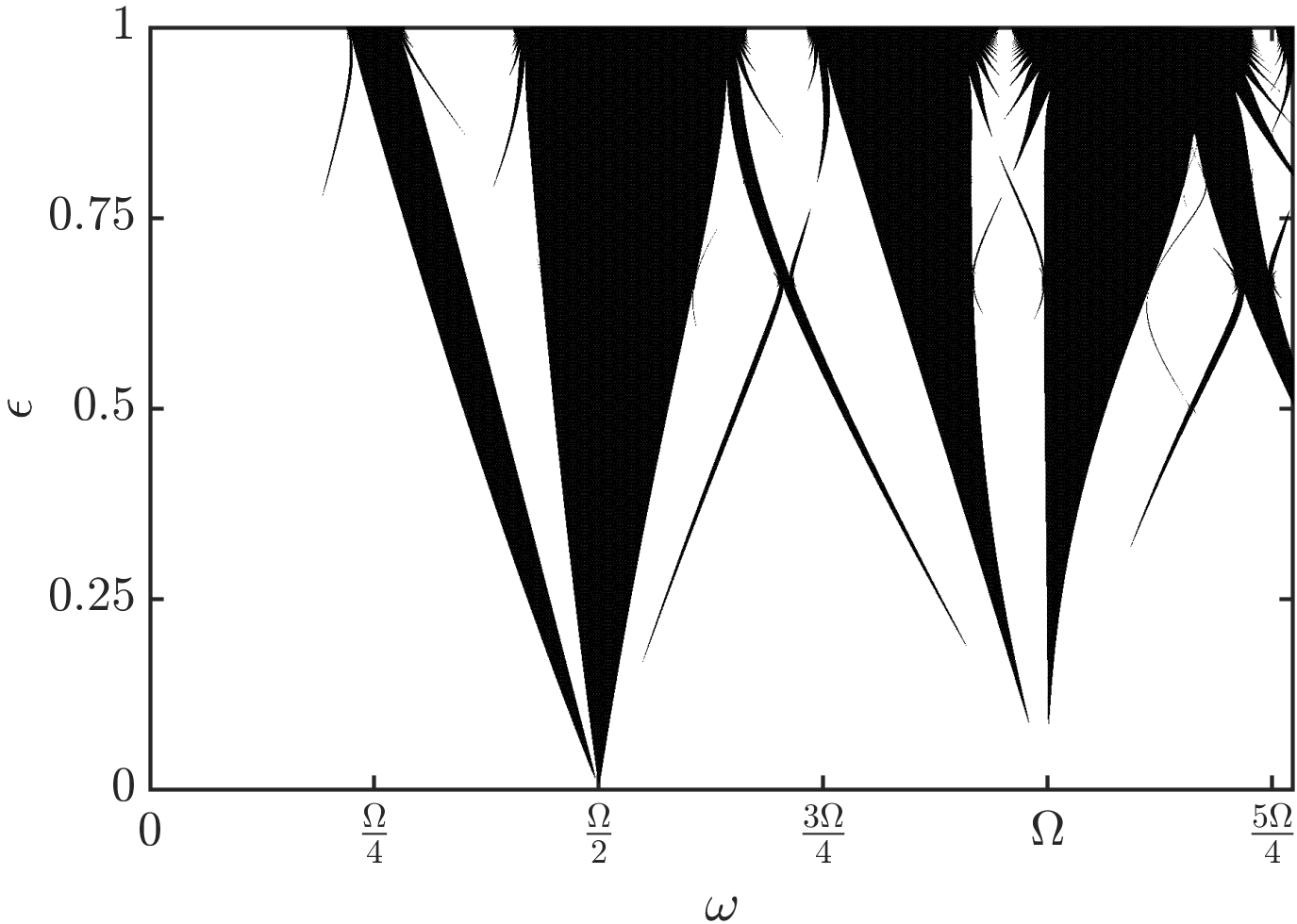}
			\caption{Numerical solution of  Eq.~(2.7) with $\mu =0.5$, $\Delta=-0.5$, $\Omega = 2\pi$ in the $\epsilon-\omega$ parameter space. Black points are the unstable solution and white region corresponds to stable solution of the \textit{Mathieu} equation. The instability zones of $V$ obtained from Eq.~(\ref{eqn:2.13}) with the same set of parameters mentioned above yield same region of the above figure.}\label{fig7}
		\end{minipage}
	\end{figure*}

	\begin{figure*}
		\begin{minipage}[b]{0.479\textwidth}
			\includegraphics[width= \textwidth]{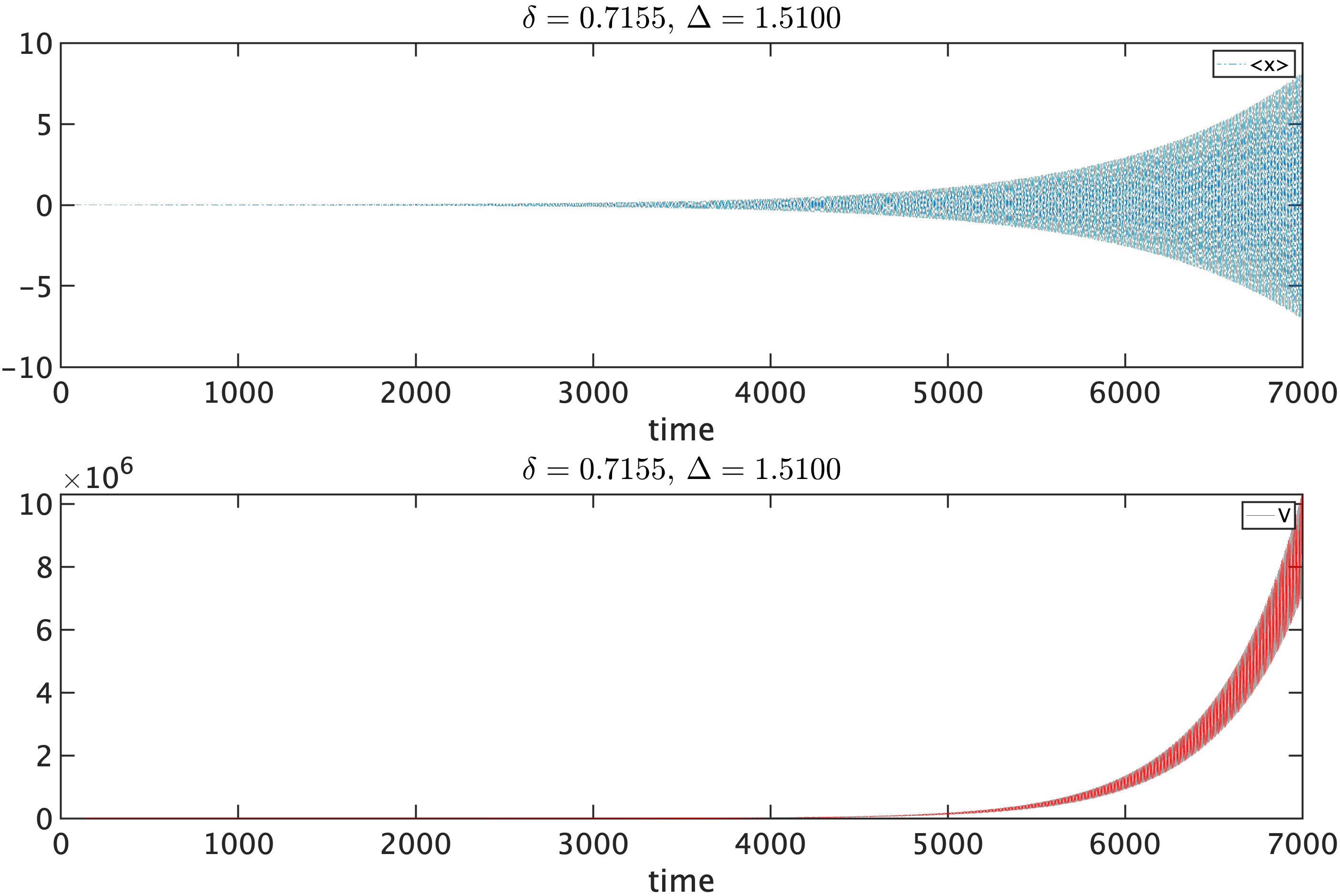}
		\end{minipage}\quad
		\begin{minipage}[b]{0.489\textwidth}
			\includegraphics[width= \textwidth]{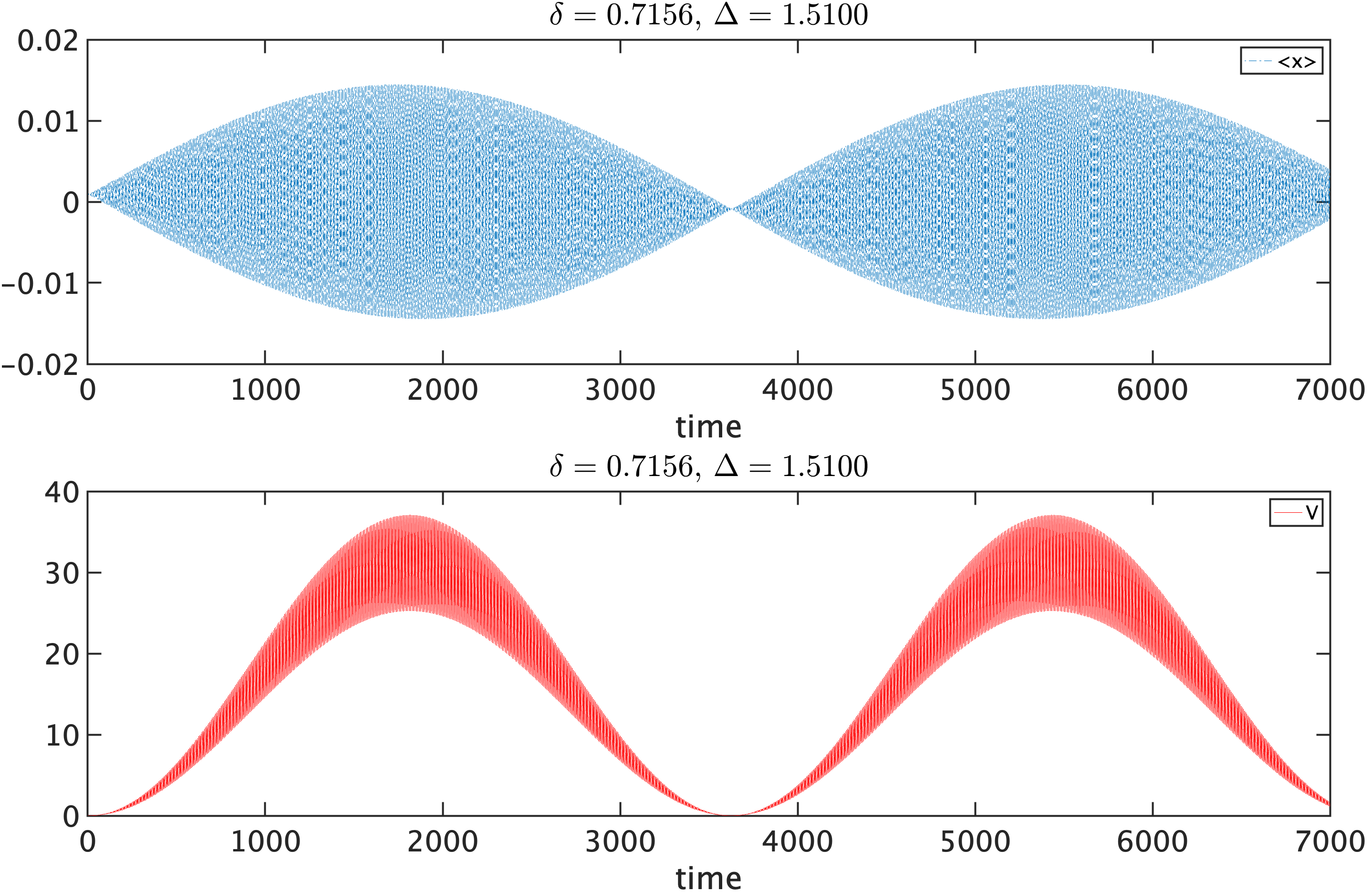}
		\end{minipage}
		\caption{Numerical solution of Eqs.~(4.3) and Eqs.~(4.12) with $\mu =1$, $\epsilon =0.1$, $\Omega = 2\pi$ and for fixed value of $\Delta =1.51$. Initial values of the corresponding variables in the equations have taken all 0.001. For $\delta =0.7155$ solutions are diverging for both slow flow of \textit{Mathieu} and variance equation, whereas for $\delta =0.7156$ both solutions are oscillatory.}\label{fig8}
	\end{figure*}
	
	Instability zones for $\mu = 0.5$ for the variance $V$ from Eqs.~(\ref{eqn:4.12a})-(\ref{eqn:4.12c}) and the exact equation (Eq.~(\ref{eqn:2.13})) are shown in Fig.~(\ref{fig5}). The similarity with Fig.~(\ref{fig4}) should be noted. We can also treat the above set of equations perturbatively as follows.
	
	In the absence of quasi-periodicity, we set $\mu=0$ and we can explore the nature of $A$,$B$ and $C$ by solving the constant coefficient linear system

	\begin{subequations}
		\begin{eqnarray}\label{eqn:4.13}
		\Delta \dot{C}&=& - \dfrac{B}{4} .\\
		\Delta \dot{A} &=& ~~~\dfrac{2 \delta_1 B}{\Omega} .\\
		\Delta \dot{B} &=&  -\dfrac{2 \delta_1 A}{\Omega}- \dfrac{C}{4} . 
		\end{eqnarray}
	\end{subequations} 
	
	The solutions are of the form $e^{\lambda t}$ with $\lambda$ obtained from 
	
	\begin{eqnarray}\label{eqn:4.14}
	\hbox {Det}\mbox{\( 
		\left(
		\begin{array}{ccc}
		\Delta\lambda &0  & \dfrac{1 }{4 }\\
		0 &    \Delta \lambda& -\dfrac{2 \delta_1 }{\Omega }\\
		\dfrac{1 }{4 } & \dfrac{2 \delta_1 }{\Omega } & \Delta \lambda  \\
		\end{array}
		\right)=0,
		\)}
	\end{eqnarray}
		
	leading to 
	
	\begin{eqnarray}\label{eqn:4.15}
	\lambda \left[(\Delta \lambda)^2  + \left(\frac{2 \delta_1 }{\Omega } \right)^2\right] - \dfrac{\lambda}{16} = 0.
	\end{eqnarray} 
	
	The three roots are found as $\Delta \lambda = 0$ and $(\Delta \lambda)^2+\dfrac{4 \delta_1 ^2 }{\Omega^2 }=\dfrac{1}{16} $. The latter yields $\Delta \lambda=\pm \sqrt{\dfrac{1}{16} - \dfrac{4 \delta_1 ^2 }{\Omega^2 }}$, and we have real $\lambda$(instability) for $|\dfrac{\delta_1}{\Omega}|< \dfrac{1}{8}$. Thus in the zone $\dfrac{\Omega}{2}+\dfrac{\epsilon \Omega}{8}$ to $\dfrac{\Omega}{2}-\dfrac{\epsilon \Omega}{8}$, we have diverging solutions of the variance $V$ which is exactly the same range for the divergence of $\langle x \rangle$ for $\mu = 0$. For $\mu \neq 0$, the divergence zone of $V$ changes and we want to find the divergence of Eqs.~(\ref{eqn:4.12a})-(\ref{eqn:4.12c}) first numerically and then perturbatively. The results as before are best presented with $\delta_1$ along the $x-$axis and $\Delta$ ( the quasi-periodicity causing ``de-tuning'' parameter) along the $y-$axis. The instability zone for $\mu = 0.5$ is shown in Fig.~(\ref{fig5}) and its striking similarity with Fig.~(\ref{fig2}) should be noted. We show the instability zones for $\langle x \rangle $ and $V$ on the same plot in Figs.~(\ref{fig1})-(\ref{fig3}) to emphasize that the zones are identical.  In Figs.~(\ref{fig6}) and (\ref{fig7})  we show the instability zones for $ \langle x \rangle$ and $V$ in the $\epsilon -\omega$ plane to show that the instability zones for $\langle x \rangle$ and $V$ are identical in the $\Delta $ vs. $\delta$ plane and in the $\epsilon$ vs. $\omega$ plane for different values of $\mu$.
	\begin{figure*}
		\centering
		\begin{minipage}[b]{0.489\textwidth}
			\includegraphics[width= \textwidth]{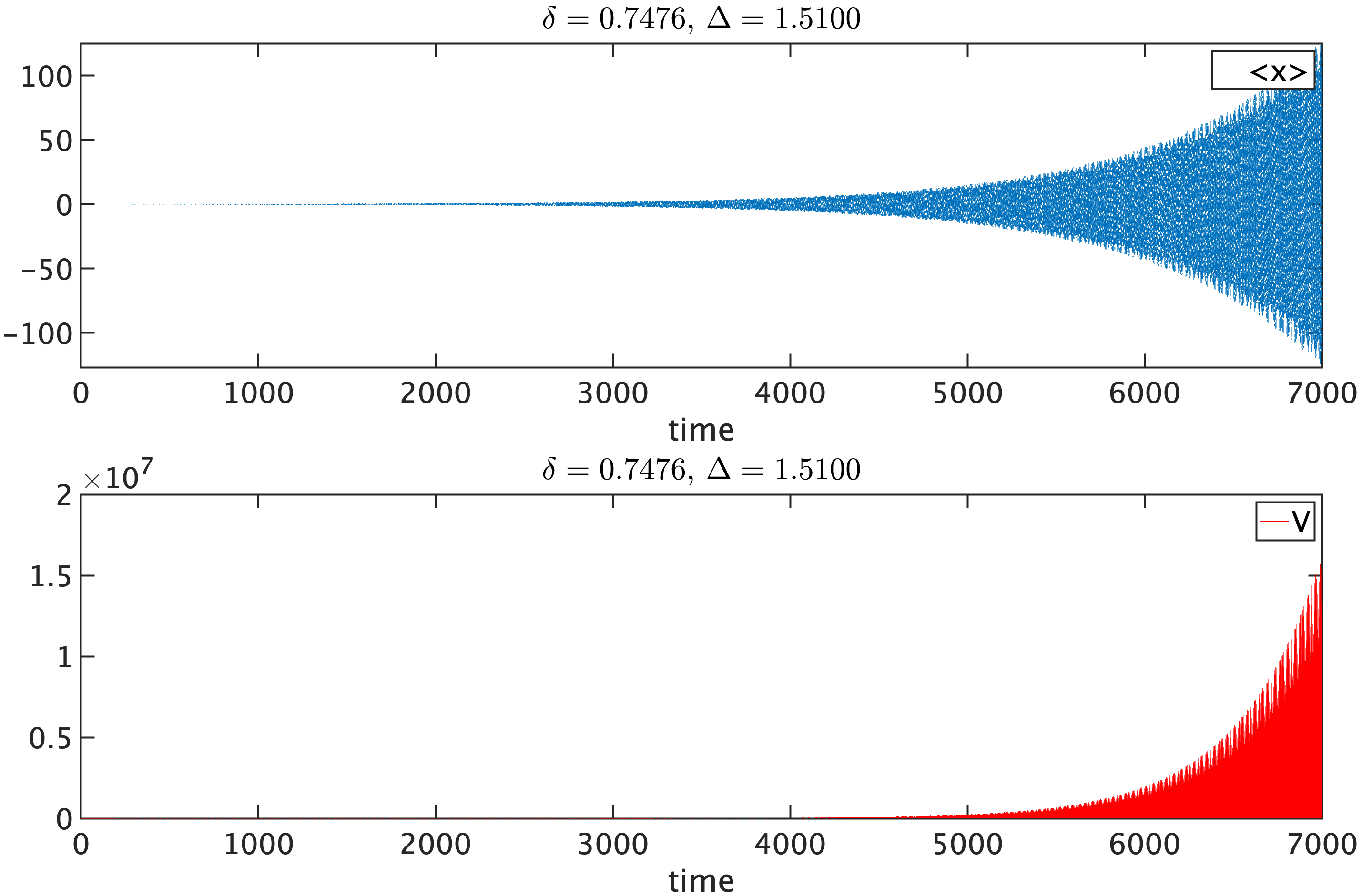}
		\end{minipage}\quad
		\begin{minipage}[b]{0.489\textwidth}
			\includegraphics[width= \textwidth]{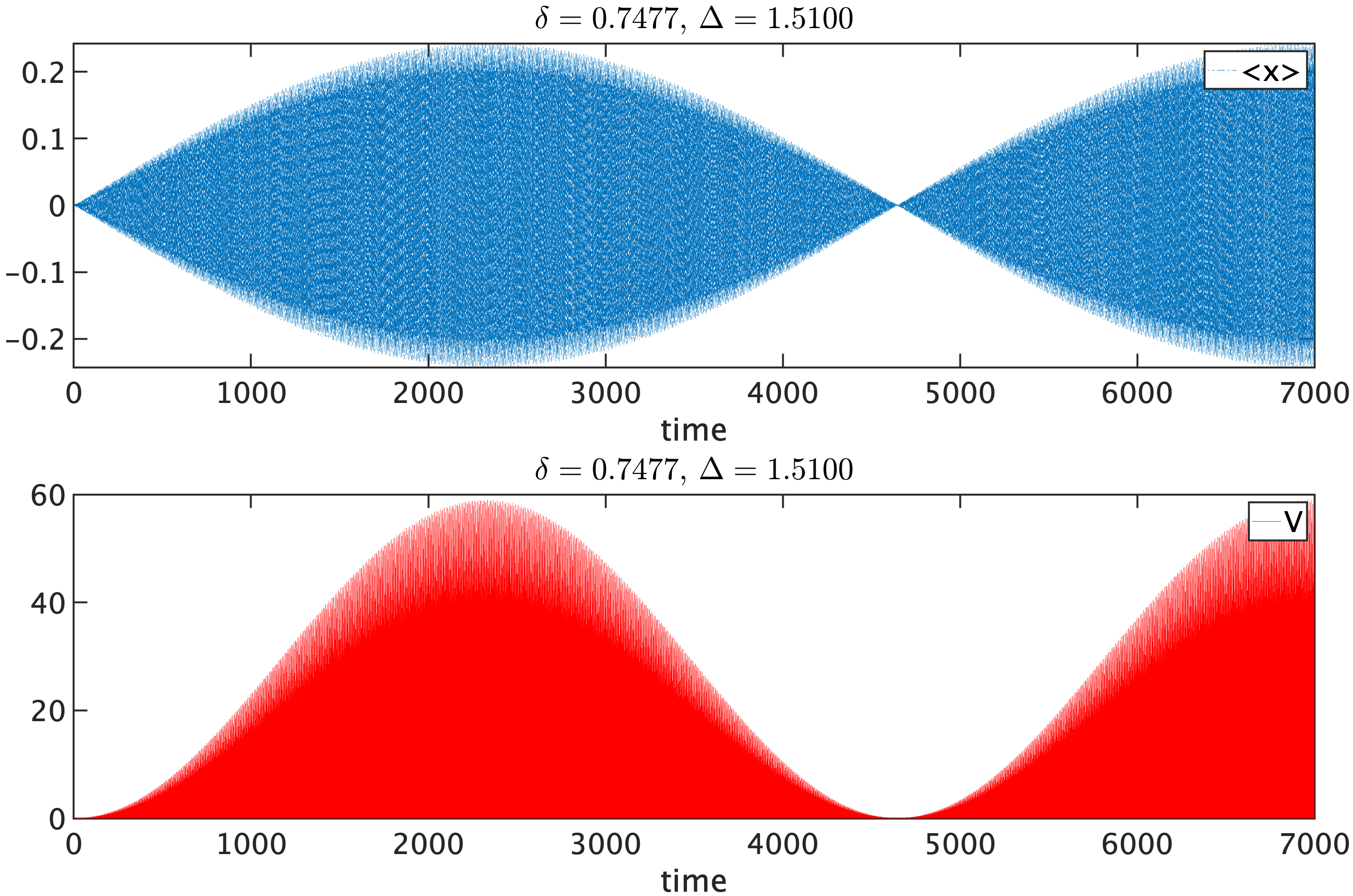}
		\end{minipage}
		\caption{Numerical solution of Eq.~(\ref{eqn:2.7}) and Eq.~(\ref{eqn:2.13}) with $\mu =1$, $\epsilon =0.1$, $\Omega = 2\pi$ and for fixed value of $\Delta =\sqrt{2}+0.1$. Initial values of the corresponding variables in the equations have taken all 0.001. For $\delta =0.7476$ solutions are diverging for both \textit{Mathieu} and variance, whereas for $\delta =0.7477$ both solutions are oscillatory.}\label{fig9}
	\end{figure*}
	
	We now turn to a perturbative treatment of the slow-flow equations (Eqs.~(\ref{eqn:4.12a})-(\ref{eqn:4.12c})) in the same manner as done for the slow flows of Eqs.~(\ref{eqn:4.3a})-(\ref{eqn:4.3b}). Accordingly we write
	
	\begin{subequations}
		\begin{eqnarray}\label{eqn:4.16a}
		A(\tau) &=& \alpha_1(\tau) \cos (\tau) + \beta_1(\tau) \sin(\tau) +\gamma_1 (\tau).\\\label{eqn:4.16b}
		B(\tau) &=&  \alpha_2(\tau) \cos (\tau) + \beta_2(\tau) \sin(\tau) +\gamma_2 (\tau).\\\label{eqn:4.16c}
		C(\tau) &=&  \alpha_3(\tau) \cos (\tau) + \beta_3(\tau) \sin(\tau) +\gamma_3 (\tau).
		\end{eqnarray}
	\end{subequations} 
	
	Inserting the equations Eqs.~(\ref{eqn:4.16a})-(\ref{eqn:4.16c}) in Eqs.~(\ref{eqn:4.12a})-(\ref{eqn:4.12c}) and equating coefficients of $1$, $\cos(\tau)$ and $\sin(\tau)$ we get a $9 \times 9$ matrix system of first order linear differential equations and the instability boundaries are found from the condition for zero eigenvalues. For $\mu =0$, the boundaries reduce exactly to those given in Eq.~(\ref{eqn:4.6}). For  $\mu \neq 0$, the system cannot be solved and does not shed any additional light implying that a perturbative treatment of the slow flow equations are not always useful! 
	
	To provide further evidence that the boundaries for instability zones of the mean and variance are identical we show the time series of the slow flow for the mean (Eqs.~ (4.3)) and the variance (Eqs.~ (4.12)) in Fig.~(\ref{fig8}). This figure clearly shows that at a fixed $\mu, \epsilon$ and $\Delta$ a change of $\delta$ from $0.7155$ to $0.7156$ changes the time series from diverging to oscillatory for both the mean and the variance. This exercise has been repeated for a variety of other parameter values. Finally, we need to address the question of whether the slow flow answers agree with the exact answers that can be obtained from Eq.~(\ref{eqn:2.7}) and (\ref{eqn:2.13}). To this end, we show a typical situation in Fig.~(\ref{fig9}). For a given $\mu, \epsilon$ and $\Delta$ it seems that change of $\delta$ from $0.7476$ to $0.7477$ changes the stability characteristic of both the mean and the variance. That this situation is always obtained is shown in Fig.~(\ref{fig4}) (mean) and Fig.~(\ref{fig5}) (variance).

	\renewcommand\thesection{\Roman{section}}
	\section{Conclusion}
	
	We have considered a quasiperiodically driven quantum parametric oscillator near the  $2:2:1$ resonance which produces the most complicated instability pattern in the three-dimensional space spanned by $\epsilon$, $\Delta$ and $\mu$. Here $\epsilon$ and $\epsilon \mu$ are the amplitudes of the quasiperiodic drives and $\Delta$ is the detuning parameter for the quasi-periodicity. We find that in the entire three-dimensional space, the instability zones of the mean position and the variance remain the same as they do in the two dimensional (frequency and amplitude) space of the periodically driven system. We establish this from perturbation theory and exact numerical solution.


\begin{thebibliography}{100} 
		\bibitem{leibfried}Leibfried, D., Blatt, R., Monroe, C., and Wineland, D.: Quantum dynamics of single trapped ions. \textsl{Reviews of Modern Physics}, \textbf{75(1)}, 281(2003).
		\bibitem{goldman}Goldman, N., and  Dalibard, J.: Periodically driven quantum systems: effective Hamiltonians and engineered gauge fields. \textsl{Physical review X}, \textbf{4(3)}, 031027(2014).
		\bibitem{chu}Chu, S. I., and Telnov, D. A.: Beyond the Floquet theorem: generalized Floquet formalisms and quasienergy methods for atomic and molecular multiphoton processes in intense laser fields. \textsl{Physics reports}, \textbf{390(1-2)}, 1-131(2004).    
		\bibitem{biswas2018}Biswas, S., Chattopadhyay, R. and Bhattacharjee, J.K.: Propagation of arbitrary initial wave packets in a quantum parametric oscillator: Instability zones for higher order moments. \textsl{Physics Letters A}. \textbf{382(18)}, 1202 (2018).
		\bibitem{biswas2019}Biswas, S., and Bhattacharjee, J. K.: On the properties of a class of higher-order Mathieu equations originating from a parametric quantum oscillator. \textsl{Nonlinear Dynamics}, \textbf{96(1)}, 737(2019).
		\bibitem{grubelnik}Grubelnik, V., Logar, M., Robnik, M., and Xia, Y.: Analysis of the parametrically periodically driven classical and quantum linear oscillator. \textsl{Physical Review E}. \textbf{99(2)}, 022209 (2019).
		\bibitem{verdeny}Verdeny, A., Puig, J., and Mintert, F.: Quasi-periodically driven quantum systems. \textsl{Z. Naturforsch}, \textbf{71(10)}, 897(2016).
		\bibitem{cubero}Cubero, D., and Renzoni, F.: Asymptotic theory of quasiperiodically driven quantum systems. \textsl{Physical Review E}. \textbf{97(6)}, 062139 (2018).
		\bibitem{crowley}Crowley, P. J., Martin, I., and Chandran, A.: Topological classification of quasiperiodically driven quantum systems. \textsl{Physical Review B}. \textbf{99(6)}, 064306(2019).
		\bibitem{jorba}Jorba, {\`A}., and Sim{\'o}, C.: On the reducibility of linear differential equations with quasiperiodic coefficients. \textsl{Journal of Differential Equations}. \textbf{98(1)}, 111 (1992).    
		
		\bibitem{krikorian}Krikorian, R.: Global density of reducible quasi-periodic cocycles on T1$\times$SU (2). \textsl{Annals of Mathematics}. \textbf{154}, 269 (2001).    
		\bibitem{avila}Avila, A., and Krikorian, R.: Reducibility or nonuniform hyperbolicity for quasiperiodic Schrödinger cocycles. \textsl{Annals of Mathematics}. \textbf{164}, 911 (2006).
		\bibitem{ballentine}Ballentine, L. E., Yang, Y., and Zibin, J. P. : Inadequacy of Ehrenfest’s theorem to characterize the classical regime. \textsl{Physical review A}. \textbf{50(4)}, 2854(1994).
		\bibitem{rand}Rand, R., Guennoun, K., and  Belhaq, M.: 2: 2: 1 Resonance in the quasiperiodic Mathieu equation. \textsl{Nonlinear Dynamics}. \textbf{31(4)}, 367 (2003).
		\bibitem{randbook}Rand, R., Zounes, R., and Hastings, R.: `A Quasiperiodic Mathieu Equation' in "Ardeshir, Guran, ed. Nonlinear Dynamics: The Richard Rand 50th Anniversary volume." \textbf{vol. 2.}, \textsl{World Scientific}. (1997).
		\bibitem{shayak}Shayak, B., and Vyas, P.: Krylov Bogoliubov type analysis of variants of the Mathieu equation. \textsl{Journal of Applied Nonlinear Dynamics}. \textbf{6(1)}, 57(2017).
		\bibitem{kovacic}Kovacic, I., Rand, R., and Sah, S. M.: Mathieu's Equation and Its Generalizations: Overview of Stability Charts and Their Features. \textsl{Applied Mechanics Reviews}. \textbf{70(2)}, 020802(2018).
	\end{thebibliography}
\end{document}